\documentclass[10pt,a4paper,onecolumn]{article}
\usepackage{marginnote}
\usepackage{graphicx}
\usepackage{xcolor}
\usepackage{authblk,etoolbox}
\usepackage{titlesec}
\usepackage{calc}
\usepackage{tikz}
\usepackage{hyperref}
\hypersetup{colorlinks,breaklinks=true,
            urlcolor=[rgb]{0.0, 0.5, 1.0},
            linkcolor=[rgb]{0.0, 0.5, 1.0}}
\usepackage{caption}
\usepackage{tcolorbox}
\usepackage{amssymb,amsmath}
\usepackage{ifxetex,ifluatex}
\usepackage{seqsplit}
\usepackage{xstring}

\usepackage{float}
\let\origfigure\figure
\let\endorigfigure\endfigure
\renewenvironment{figure}[1][2] {
    \expandafter\origfigure\expandafter[H]
} {
    \endorigfigure
}

\usepackage{fixltx2e} 
\usepackage[
  backend=biber,
]{biblatex}
\bibliography{paper.bib}

\let\textttOrig=\texttt
\def\texttt#1{\expandafter\textttOrig{\seqsplit{#1}}}
\renewcommand{\seqinsert}{\ifmmode
  \allowbreak
  \else\penalty6000\hspace{0pt plus 0.02em}\fi}

\makeatletter
\let\href@Orig=\href
\def\href@Urllike#1#2{\href@Orig{#1}{\begingroup
    \def\Url@String{#2}\Url@FormatString
    \endgroup}}
\def\href@Notdoi#1#2{\def\tempa{#1}\def\tempb{#2}%
  \ifx\tempa\tempb\relax\href@Urllike{#1}{#2}\else
  \href@Orig{#1}{#2}\fi}
\def\href#1#2{%
  \IfBeginWith{#1}{https://doi.org}%
  {\href@Urllike{#1}{#2}}{\href@Notdoi{#1}{#2}}}
\makeatother

\newlength{\cslhangindent}
\newlength{\csllabelwidth}
\newenvironment{CSLReferences}[3] 
 {
  \setlength{\parindent}{0pt}
  \ifodd #1 \everypar{\setlength{\hangindent}{\cslhangindent}}\ignorespaces\fi
  \ifnum #2 > 0
  \setlength{\parskip}{#2\baselineskip}
  \fi
 }%
 {}
\usepackage{calc}

\usepackage[top=3.5cm, bottom=3cm, right=1.5cm, left=1.0cm,
            headheight=2.2cm, reversemp, includemp, marginparwidth=4.5cm]{geometry}

\titleformat{\section}
  {\normalfont\sffamily\Large\bfseries}
  {}{0pt}{}
\titleformat{\subsection}
  {\normalfont\sffamily\large\bfseries}
  {}{0pt}{}
\titleformat{\subsubsection}
  {\normalfont\sffamily\bfseries}
  {}{0pt}{}
\titleformat*{\paragraph}
  {\sffamily\normalsize}

\usepackage{fancyhdr}
\makeatletter
\let\ps@plain\ps@fancy
\fancyheadoffset[L]{4.5cm}
\fancyfootoffset[L]{4.5cm}

\definecolor{linky}{rgb}{0.0, 0.5, 1.0}

\newtcolorbox{repobox}
   {colback=red, colframe=red!75!black,
     boxrule=0.5pt, arc=2pt, left=6pt, right=6pt, top=3pt, bottom=3pt}

\newcommand{\ExternalLink}{%
   \tikz[x=1.2ex, y=1.2ex, baseline=-0.05ex]{%
       \begin{scope}[x=1ex, y=1ex]
           \clip (-0.1,-0.1)
               --++ (-0, 1.2)
               --++ (0.6, 0)
               --++ (0, -0.6)
               --++ (0.6, 0)
               --++ (0, -1);
           \path[draw,
               line width = 0.5,
               rounded corners=0.5]
               (0,0) rectangle (1,1);
       \end{scope}
       \path[draw, line width = 0.5] (0.5, 0.5)
           -- (1, 1);
       \path[draw, line width = 0.5] (0.6, 1)
           -- (1, 1) -- (1, 0.6);
       }
   }

\patchcmd{\@maketitle}{center}{flushleft}{}{}
\patchcmd{\@maketitle}{center}{flushleft}{}{}
\patchcmd{\@maketitle}{\LARGE}{\LARGE\sffamily}{}{}
\def\maketitle{{%
  
  \AB@maketitle}}
\makeatletter
\renewcommand\AB@affilsepx{ \protect\Affilfont}
\renewcommand\AB@affilnote[1]{{\bfseries #1}\hspace{3pt}}
\renewcommand{\affil}[2][]%
   {\newaffiltrue\let\AB@blk@and\AB@pand
      \if\relax#1\relax\def\AB@note{\AB@thenote}\else\def\AB@note{#1}%
        \setcounter{Maxaffil}{0}\fi
        \begingroup
        \let\href=\href@Orig
        \let\texttt=\textttOrig
        \let\protect\@unexpandable@protect
        \def\thanks{\protect\thanks}\def\footnote{\protect\footnote}%
        \@temptokena=\expandafter{\AB@authors}%
        {\def\\{\protect\\\protect\Affilfont}\xdef\AB@temp{#2}}%
         \xdef\AB@authors{\the\@temptokena\AB@las\AB@au@str
         \protect\\[\affilsep]\protect\Affilfont\AB@temp}%
         \gdef\AB@las{}\gdef\AB@au@str{}%
        {\def\\{, \ignorespaces}\xdef\AB@temp{#2}}%
        \@temptokena=\expandafter{\AB@affillist}%
        \xdef\AB@affillist{\the\@temptokena \AB@affilsep
          \AB@affilnote{\AB@note}\protect\Affilfont\AB@temp}%
      \endgroup
       \let\AB@affilsep\AB@affilsepx
}
\makeatother

\renewcommand\Affilfont{\sffamily\small\mdseries}
\ifnum 0\ifxetex 1\fi\ifluatex 1\fi=0 
  \usepackage[T1]{fontenc}
  \usepackage[utf8]{inputenc}

\else 
  \ifxetex
    \usepackage{mathspec}
    \usepackage{fontspec}

  \else
    \usepackage{fontspec}
  \fi
  \defaultfontfeatures{Ligatures=TeX,Scale=MatchLowercase}

\fi
\IfFileExists{upquote.sty}{\usepackage{upquote}}{}
\IfFileExists{microtype.sty}{%
\usepackage{microtype}
\UseMicrotypeSet[protrusion]{basicmath} 
}{}

\usepackage{hyperref}
\hypersetup{unicode=true,
            pdftitle={The Python Sky Model 3 software},
            pdfborder={0 0 0},
            breaklinks=true}
\usepackage{longtable,booktabs}

\let\addcontentslineOrig=\addcontentsline
\def\addcontentsline#1#2#3{\bgroup
  \let\texttt=\textttOrig\addcontentslineOrig{#1}{#2}{#3}\egroup}
\let\markbothOrig\markboth
\def\markboth#1#2{\bgroup
  \let\texttt=\textttOrig\markbothOrig{#1}{#2}\egroup}
\let\markrightOrig\markright
\def\markright#1{\bgroup
  \let\texttt=\textttOrig\markrightOrig{#1}\egroup}

\IfFileExists{parskip.sty}{%
\usepackage{parskip}
}{
\setlength{\parindent}{0pt}
\setlength{\parskip}{6pt plus 2pt minus 1pt}
}
\providecommand{\tightlist}{%
  \setlength{\itemsep}{0pt}\setlength{\parskip}{0pt}}
\ifx\paragraph\undefined\else
\let\oldparagraph\paragraph
\renewcommand{\paragraph}[1]{\oldparagraph{#1}\mbox{}}
\fi
\ifx\subparagraph\undefined\else
\let\oldsubparagraph\subparagraph
\renewcommand{\subparagraph}[1]{\oldsubparagraph{#1}\mbox{}}
\fi

\title{The Python Sky Model 3 software}

        \author[1]{Andrea Zonca}
          \author[2]{Ben Thorne}
          \author[3,4,5]{Nicoletta Krachmalnicoff}
          \author[6,7]{Julian Borrill}
    
      \affil[1]{San Diego Supercomputer Center, University of California
San Diego, San Diego, USA}
      \affil[2]{Department of Physics, University of California Davis,
One Shields Avenue, Davis, CA 95616, USA}
      \affil[3]{SISSA, Via Bonomea 265, 34136 Trieste, Italy}
      \affil[4]{INFN, Via Valerio 2, 34127 Trieste, Italy}
      \affil[5]{IFPU, Via Beirut 2, 34014 Trieste, Italy}
      \affil[6]{Computational Cosmology Center, Lawrence Berkeley
National Laboratory, Berkeley, CA 94720, USA}
      \affil[7]{Space Sciences Laboratory at University of California, 7
Gauss Way, Berkeley, CA 94720}
  \date{\vspace{-7ex}}

\begin{document}
\maketitle

\marginpar{

  \begin{flushleft}
  \sffamily\small

  {\bfseries DOI:} \href{https://doi.org/DOI unavailable}{\color{linky}{DOI unavailable}}

  \vspace{2mm}

  {\bfseries Software}
  \begin{itemize}
    \setlength\itemsep{0em}
    \item \href{N/A}{\color{linky}{Review}} \ExternalLink
    \item \href{NO_REPOSITORY}{\color{linky}{Repository}} \ExternalLink
    \item \href{DOI unavailable}{\color{linky}{Archive}} \ExternalLink
  \end{itemize}

  \vspace{2mm}

  \par\noindent\hrulefill\par

  \vspace{2mm}

  {\bfseries Editor:} \href{https://example.com}{Pending
Editor} \ExternalLink \\
  \vspace{1mm}
    {\bfseries Reviewers:}
  \begin{itemize}
  \setlength\itemsep{0em}
    \item \href{https://github.com/Pending Reviewers}{@Pending
Reviewers}
    \end{itemize}
    \vspace{2mm}

  {\bfseries Submitted:} N/A\\
  {\bfseries Published:} N/A

  \vspace{2mm}
  {\bfseries License}\\
  Authors of papers retain copyright and release the work under a Creative Commons Attribution 4.0 International License (\href{http://creativecommons.org/licenses/by/4.0/}{\color{linky}{CC BY 4.0}}).

  \end{flushleft}
}

\hypertarget{statement-of-need}{%
\section{Statement of Need}\label{statement-of-need}}

The Cosmic Microwave Background (CMB) radiation, emitted just 370
thousand years after the Big Bang, is a pristine probe of the Early
Universe. After being emitted at high temperatures, the CMB was
redshifted by the subsequent 13.8 billion years of cosmic expansion,
such that it is brightest at microwave frequencies today. However, our
own Milky Way galaxy also emits in the microwave portion of the
spectrum, obscuring our view of the CMB. Examples of this emission are
thermal radiation by interstellar dust grains, and synchrotron emission
by relativistic electrons spiraling in magnetic fields. Cosmologists
need to create synthetic maps of the CMB and of the galactic emission
based on available data and on physical models that extrapolate
observations to different frequencies. The resulting maps are useful to
test data reduction algorithms, to understand residual systematics, to
forecast maps produced by future instruments, to run Monte Carlo
analysis for noise estimation, and more.

\hypertarget{summary}{%
\section{Summary}\label{summary}}

The Python Sky Model (PySM) is a Python package used by Cosmic Microwave
Background (CMB) experiments to simulate maps, in HEALPix (Górski et
al., 2005; Zonca et al., 2019) pixelization, of the various diffuse
astrophysical components of Galactic emission relevant at CMB
frequencies (i.e.~dust, synchrotron, free-free and Anomalous Microwave
Emission), as well as the CMB itself. These maps may be integrated over
a given instrument bandpass and smoothed with a given instrument beam.
The template emission maps used by PySM are based on Planck (Planck
Collaboration, 2018) and WMAP (Bennett et al., 2013) data and are
noise-dominated at small scales. Therefore, PySM simulation templates
are smoothed to retain the large-scale information, and then
supplemented with modulated Gaussian realizations at smaller scales.
This strategy allows one to simulate data at higher resolution than the
input maps.

PySM 2 (Thorne et al., 2017), released in 2016, has become the de-facto
standard for simulating Galactic emission, for example it is used by
CMB-S4, Simons Observatory, LiteBird, PICO, CLASS, POLARBEAR and other
CMB experiments, as shown by the
\href{https://scholar.google.com/scholar?start=0\&hl=en\&as_sdt=2005\&sciodt=0,5\&cites=16628417670342266167\&scipsc=}{80+
citations of the PySM 2 publication}. As the resolution of upcoming
experiments increases, the PySM 2 software has started to show some
limitations:

\begin{itemize}
\tightlist
\item
  Emission templates are provided at 7.9 arcminutes resolution (HEALPix
  \(N_{side}=512\)), while the next generation of CMB experiments will
  require sub-arcminute resolution.
\item
  The software is implemented in pure \texttt{numpy}, meaning that it
  has significant memory overhead and is not multi-threaded, precluding
  simply replacing the current templates with higher-resolution versions
\item
  Emission templates are included in the PySM 2 Python package, this is
  still practical when each of the roughly 40 input maps is
  \textasciitilde10 Megabytes, but will not be if they are over 1
  Gigabyte.
\end{itemize}

The solution to these issues was to reimplement PySM from scratch
focusing on these features:

\begin{itemize}
\tightlist
\item
  Reimplement all the models with the \texttt{numba} (Lam et al., 2015)
  Just-In-Time compiler for Python to reduce memory overhead and
  optimize performance: the whole integration loop of a template map
  over the frequency response of an instrument is performed in a single
  pass in automatically compiled and multi-threaded Python code.
\item
  Use MPI through \texttt{mpi4py} to coordinate execution of PySM 3
  across multiple nodes, this allows to support template maps at a
  resolution up to 0.4 arcminutes (HEALPix \(N_{side}=8192\)).
\item
  Rely on \texttt{libsharp} (Reinecke \& Seljebotn, 2013), a distributed
  implementation of spherical harmonic transforms, to smooth the maps
  with the instrument beam when maps are distributed over multiple nodes
  with MPI.
\item
  Employ the data utilities infrastructure provided by \texttt{astropy}
  (Astropy Collaboration et al., 2018, 2013) to download the input
  templates and cache them when requested.
\end{itemize}

At this stage we strive to maintain full compatibility with PySM 2,
therefore we implement the exact same astrophysical emission models with
the same naming scheme. In the extensive test suite we compare the
output of each PySM 3 model with the results obtained by PySM 2.

\hypertarget{performance}{%
\section{Performance}\label{performance}}

As an example of the performance improvements achieved with PySM 3 over
PySM 2, we run the following configuration:

\begin{itemize}
\tightlist
\item
  An instrument with 3 channels, with different beams, and a top-hat
  bandpass defined numerically at 10 frequency samples.
\item
  A sky model with the simplest models of dust, synchrotron, free-free
  and AME {[}\texttt{a1,d1,s1,f1} in PySM terms{]}.
\item
  Execute on a 12-core Intel processor with 12 GB of RAM.
\end{itemize}

The following tables shows the walltime and peak memory usage of this
simulation executed at the native PySM 2 resolution of \(N_{side}=512\)
and at two higher resolutions:

\begin{longtable}[]{@{}lll@{}}
\toprule
Output \(N_{side}\) & PySM 3 & PySM 2\tabularnewline
\midrule
\endhead
512 & 1m 0.7 GB & 1m40s 1.45 GB\tabularnewline
1024 & 3m30s 2.3 GB & 7m20s 5.5 GB\tabularnewline
2048 & 16m10s 8.5 GB & Out of memory\tabularnewline
\bottomrule
\end{longtable}

At the moment it is not very useful to run at resolutions higher than
\(N_{side}=512\) because there is no actual template signal at smaller
scales. However, it demonstrates the performance improvements that will
make working with higher resolution templates possible.

\hypertarget{future-work}{%
\section{Future work}\label{future-work}}

PySM 3 opens the way to implement a new category of models at much
higher resolution. However, instead of just upgrading the current models
to smaller scales we want to also update them with the latest knowledge
of Galactic emission and gather feedback from each of the numerous CMB
experiments. For this reason we are collaborating with the Panexperiment
Galactic Science group to lead the development of the new class of
models to be included in PySM 3.

\hypertarget{how-to-cite}{%
\section{How to cite}\label{how-to-cite}}

If you are using PySM 3 for your work, please cite this paper for the
software itself; for the actual emission modeling please also cite the
original PySM 2 paper (Thorne et al., 2017). There will be a future
paper on the generation of new PySM 3 astrophysical models.

\hypertarget{acknowledgments}{%
\section{Acknowledgments}\label{acknowledgments}}

\begin{itemize}
\tightlist
\item
  This work was supported in part by NASA grant \texttt{80NSSC18K1487}.
\item
  The software was tested, in part, on facilities run by the Scientific
  Computing Core of the Flatiron Institute.
\item
  This research used resources of the National Energy Research
  Scientific Computing Center (NERSC), a U.S. Department of Energy
  Office of Science User Facility located at Lawrence Berkeley National
  Laboratory, operated under Contract No.~\texttt{DE-AC02-05CH11231}.
\end{itemize}

\hypertarget{references}{%
\section*{References}\label{references}}
\addcontentsline{toc}{section}{References}

\hypertarget{refs}{}
\begin{CSLReferences}{1}{0}
\leavevmode\hypertarget{ref-astropy2018}{}%
Astropy Collaboration, Price-Whelan, A. M., Sipőcz, B. M., Günther, H.
M., Lim, P. L., Crawford, S. M., Conseil, S., Shupe, D. L., Craig, M.
W., Dencheva, N., Ginsburg, A., Vand erPlas, J. T., Bradley, L. D.,
Pérez-Suárez, D., de Val-Borro, M., Aldcroft, T. L., Cruz, K. L.,
Robitaille, T. P., Tollerud, E. J., \ldots{} Astropy Contributors.
(2018). {The Astropy Project: Building an Open-science Project and
Status of the v2.0 Core Package}. \emph{156}(3), 123.
\url{https://doi.org/10.3847/1538-3881/aabc4f}

\leavevmode\hypertarget{ref-astropy2013}{}%
Astropy Collaboration, Robitaille, T. P., Tollerud, E. J., Greenfield,
P., Droettboom, M., Bray, E., Aldcroft, T., Davis, M., Ginsburg, A.,
Price-Whelan, A. M., Kerzendorf, W. E., Conley, A., Crighton, N.,
Barbary, K., Muna, D., Ferguson, H., Grollier, F., Parikh, M. M., Nair,
P. H., \ldots{} Streicher, O. (2013). {Astropy: A community Python
package for astronomy}. \emph{558}, A33.
\url{https://doi.org/10.1051/0004-6361/201322068}

\leavevmode\hypertarget{ref-wmap13}{}%
Bennett, C. L., Larson, D., Weiland, J. L., Jarosik, N., Hinshaw, G.,
Odegard, N., Smith, K. M., Hill, R. S., Gold, B., Halpern, M., Komatsu,
E., Nolta, M. R., Page, L., Spergel, D. N., Wollack, E., Dunkley, J.,
Kogut, A., Limon, M., Meyer, S. S., \ldots{} Wright, E. L. (2013).
{Nine-year Wilkinson Microwave Anisotropy Probe (WMAP) Observations:
Final Maps and Results}. \emph{208}, 20.
\url{https://doi.org/10.1088/0067-0049/208/2/20}

\leavevmode\hypertarget{ref-gorski05}{}%
Górski, K. M., Hivon, E., Banday, A. J., Wandelt, B. D., Hansen, F. K.,
Reinecke, M., \& Bartelmann, M. (2005). {HEALPix: A Framework for
High-Resolution Discretization and Fast Analysis of Data Distributed on
the Sphere}. \emph{622}, 759--771. \url{https://doi.org/10.1086/427976}

\leavevmode\hypertarget{ref-numba}{}%
Lam, S. K., Pitrou, A., \& Seibert, S. (2015). Numba: A LLVM-based
python JIT compiler. \emph{Proceedings of the Second Workshop on the
LLVM Compiler Infrastructure in HPC}.
\url{https://doi.org/10.1145/2833157.2833162}

\leavevmode\hypertarget{ref-planck18}{}%
Planck Collaboration. (2018). \emph{{Planck 2018 results. I. Overview
and the cosmological legacy of Planck}}.
\url{http://arxiv.org/abs/1807.06205}

\leavevmode\hypertarget{ref-libsharp}{}%
Reinecke, M., \& Seljebotn, D. S. (2013). Libsharp -- spherical harmonic
transforms revisited. \emph{Astronomy \& Astrophysics}, \emph{554},
A112. \url{https://doi.org/10.1051/0004-6361/201321494}

\leavevmode\hypertarget{ref-pysm17}{}%
Thorne, B., Dunkley, J., Alonso, D., \& Næss, S. (2017). The python sky
model: Software for simulating the galactic microwave sky. \emph{Monthly
Notices of the Royal Astronomical Society}, \emph{469}(3), 2821--2833.
\url{https://doi.org/10.1093/mnras/stx949}

\leavevmode\hypertarget{ref-healpy09}{}%
Zonca, A., Singer, L. P., Lenz, D., Reinecke, M., Rosset, C., Hivon, E.,
\& Gorski, K. M. (2019). Healpy: Equal area pixelization and spherical
harmonics transforms for data on the sphere in python. \emph{Journal of
Open Source Software}, \emph{4}(35), 1298.
\url{https://doi.org/10.21105/joss.01298}

\end{CSLReferences}

\end{document}